\documentclass[a4paper,11pt]{article}
\usepackage{pos}
\usepackage{changepage}

\title{A VERITAS view of HESS~J1857+026 within a multi-wavelength analysis}

\author*[a]{Yu Chen}
\onbehalf{for the VERITAS Collaboration}
\affiliation[a]{Department of Physics and Astronomy, University of California, Los Angeles\\
  475 Portola Plaza, Los Angeles, CA 90095, USA}

\emailAdd{ychen@astro.ucla.edu}

\abstract{HESS~J1857+026 remains a mysterious gamma-ray emitter since its discovery in 2008. Despite the disclosure of a nearby pulsar and multiple studies in the high-energy (HE, E > 100 MeV) and very-high-energy (VHE, E > 100 GeV) regimes, there have been no confirmed counterparts (e.g., an SNR shell or other extended structure) in X-ray or other wavelengths. We present the result of our study of the VHE emission of HESS~J1857+026 with VERITAS as part of a multi-wavelength investigation to uncover its emission mechanisms. Our result confirms the extended nature of the source and we characterize its spectral and morphological features in the VHE band. Using the morphology of the source revealed in our analysis, we also explore the underlying transport process of a possible electron population in a leptonic PWN scenario for the gamma-ray emission.}

\ConferenceLogo{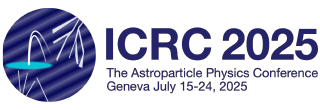}

\FullConference{39th International Cosmic Ray Conference (ICRC2025)\\
 15–24 July 2025\\
Geneva, Switzerland\\}

\begin{document}
\maketitle

\section{Introduction}
HESS~J1857+026 was first described by HESS as one of the eight newly discovered VHE sources without any counterparts in their Galactic plane survey \cite{hess_2008}. A spatially coincident pulsar PSR~J1856+0245 was later discovered with Arecibo \cite{hessels_psr_2008}. The pulsar was shown to have a characteristic age of 21\,kyr and a spin-down luminosity of $4.6\,\times\,10^{36}$ ergs s$^{-1}$ \cite{hessels_psr_2008}, which is energetic enough to power the TeV emission, and thus a pulsar wind nebula (PWN) scenario was proposed. 

Many studies were performed to investigate the nature of the source. \textit{Fermi}-LAT detected an extended emission in GeV energies with a disk extension of 0.61$^\circ$ \cite{Fermi_2017_extsources}. MAGIC showed a spatially resolved two-source scenario in energies >\,1\,TeV where they associated the southern source as a PWN driven by PSR~J1856+0245 while the northern source is related to a CO cavity and an HII region U36.40+0.02 \cite{magic_2014}. Petriella et.al. reported an HI shell enclosing the entire HESS TeV emission region in a cavity-like structure, and the kinematics of the HI shell resembles a superbubble \cite{petriella_radio_2021}, indicating a possible single-source origin. Despite the differences, most studies favor at least a partial leptonic PWN scenario because of the non-detection of atomic or molecular material in the region to account for the extended TeV emission with pure hadronic interaction.

The inhibited diffusion of electrons and positrons around pulsars has been an active field of research in the past decade. It was first discovered by HAWC that the diffusion coefficient around Geminga and PSR B0656+14 is about $10^{27}$\,cm$^2$/s at 100\,TeV \cite{HAWC_2017_Geminga}, orders of magnitude below the Galactic average \cite{genolini_cosmic-ray_2019}. The conclusion was extended to many Galactic pulsars by \cite{di_mauro_evidences_2020} where many were found to have a diffusion coefficient in the nearby region at $10^{26}$\,cm$^2$/s at 1\,TeV. Recently, HAWC identified Geminga-like halo emission around 36 middle-aged pulsars and provided further evidence that inhibited diffusion around pulsars may be quite common \cite{HAWC_2025_halo}.

This work is part of a multi-wavelength investigation on HESS~J1857+026 including \textit{Fermi}-LAT, VERITAS, and HAWC data. Here we focus on the VHE analysis with VERITAS data and provide complementary analysis details that could not fit in a future multi-wavelength paper. We will also calculate the diffusion coefficient given the morphological features of the source and show our results in Section \ref{sec:results}.

\section{VERITAS observation and analysis}

\subsection{The VERITAS instrument and observation}
VERITAS is an array of 4 imaging atmospheric Cherenkov telescopes (IACTs) located at the Fred Lawrence Whipple Observatory in southern Arizona (31$^\circ$40'N, 110$^\circ$57'W,  1.3\,km above sea level), USA. Each IACT has a reflector that is 12\,m in diameter and a field of view (FOV) of about 3.5$^\circ$ across. VERITAS is sensitive to gamma rays in the energy range of 100 GeV to above 30 TeV and has an angular resolution R$_{68}$\,<\,0.1$^\circ$ at 1\,TeV. It can detect point sources with a flux of 1\% Crab in 25\,h.

VERITAS has observed the region of HESS~J1857+026 from 2008 to 2016, including serendipitous observation of other targets, e.g., the supernova remnant W44, in the FOV. After quality selection requiring good weather and a stable trigger rate, about 30\,h of data are used in this analysis. The mean elevation of observation in the analysis dataset is 58.21$^\circ$ and the mean offset from the TeV center described by HESS (\cite{hess_2018_GPS}) is about 1.21$^\circ$. 

\subsection{VERITAS analysis}\label{sec:VTS-analysis}
The event reconstruction and gamma-hadron separation are performed with EventDisplay v490.7 \cite{maier2017eventdisplay,Maier_Eventdisplay_An_Analysis_2024}. The data are then converted to DL3 format using V2DL3\footnote{\url{https://github.com/open-gamma-ray-astro/gamma-astro-data-formats}}. Gammapy-tools\footnote{\url{https://github.com/VERITAS-Observatory/gammapy-tools}} are used to generate background templates (a.k.a. 2D acceptance maps) for each observation run that are attached to the DL3 files. More details about the background generation will be given in the upcoming multi-wavelength paper. 

The DL3 data are analyzed with Gammapy v1.3 \cite{gammapy_zenodo_v1p3,gammapy:2023}. Exclusion regions are set around PSR~J1856+0245 for a radius of 1$^\circ$, and around HESS~J1858+020 and W44 for a radius of 0.3$^\circ$. A 3D stacked analysis with a FOV background method is performed. The background rates for each run are adjusted using the `fit' method (i.e. according to a power-law distribution) during data reduction and are fixed in the following steps. We fit HESS~J1857+026 and another source HESS~J1858+020 in the FOV with Gammapy Skymodels that incorporate a power-law spectral model 
\begin{equation}
  \frac{dN}{dE} = N_{0} \left(\frac{E}{E_0}\right)^{-\Gamma}
  \label{equation:pl}
\end{equation}
where ${E_0}$ is fixed at 1\,TeV and a symmetric Gaussian spatial model. Optimized parameters are found by maximizing the likelihood of the models on the binned dataset. A fit mask with a low-energy bound of 300\,GeV (a cut on the dataset that removes lower energy bins) is also applied in the fit to remove artifacts in the sky map. 

\begin{figure}[h]
    \centering
    \includegraphics[width=0.47\linewidth]{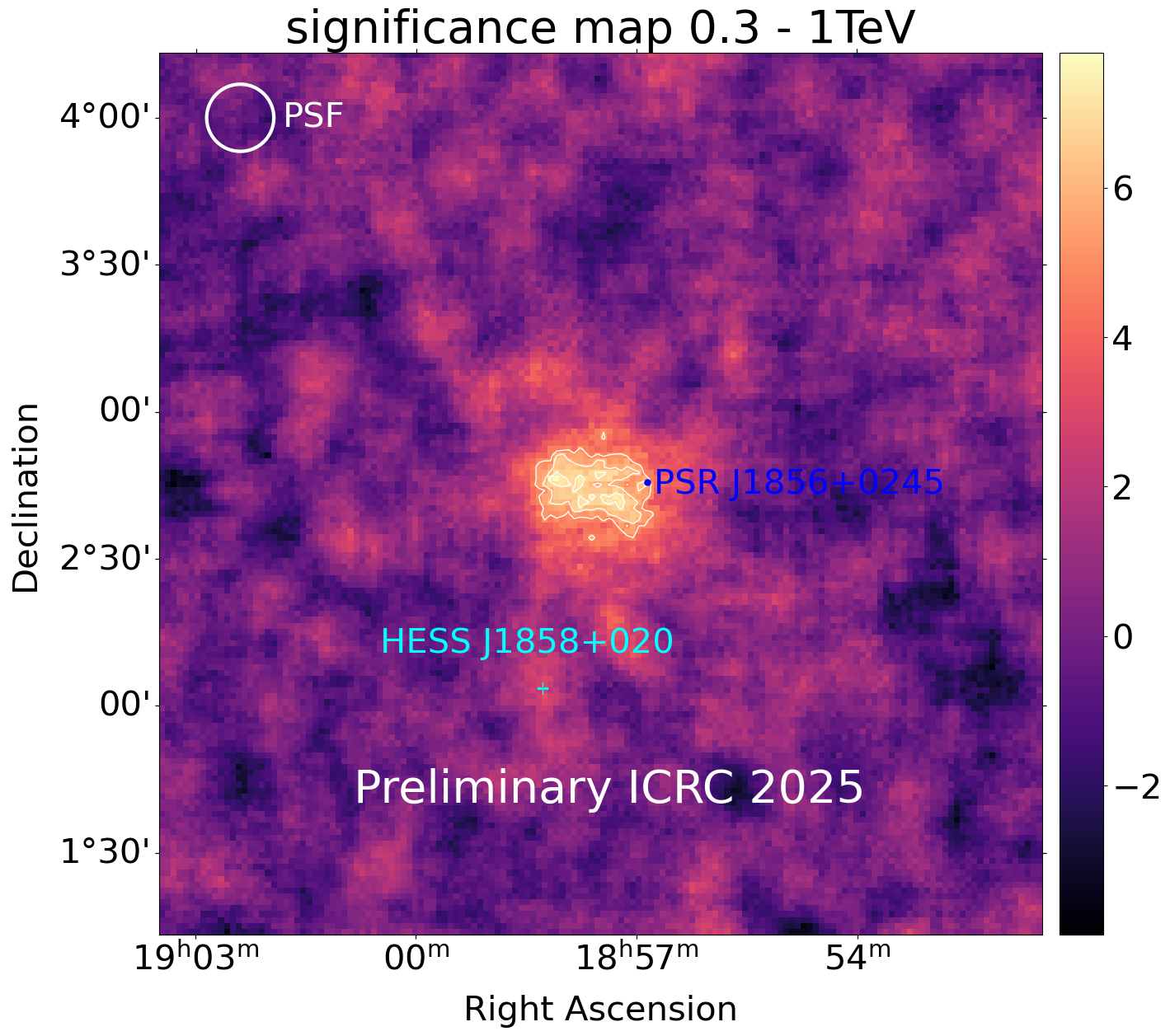}
    \includegraphics[width=0.47\linewidth]{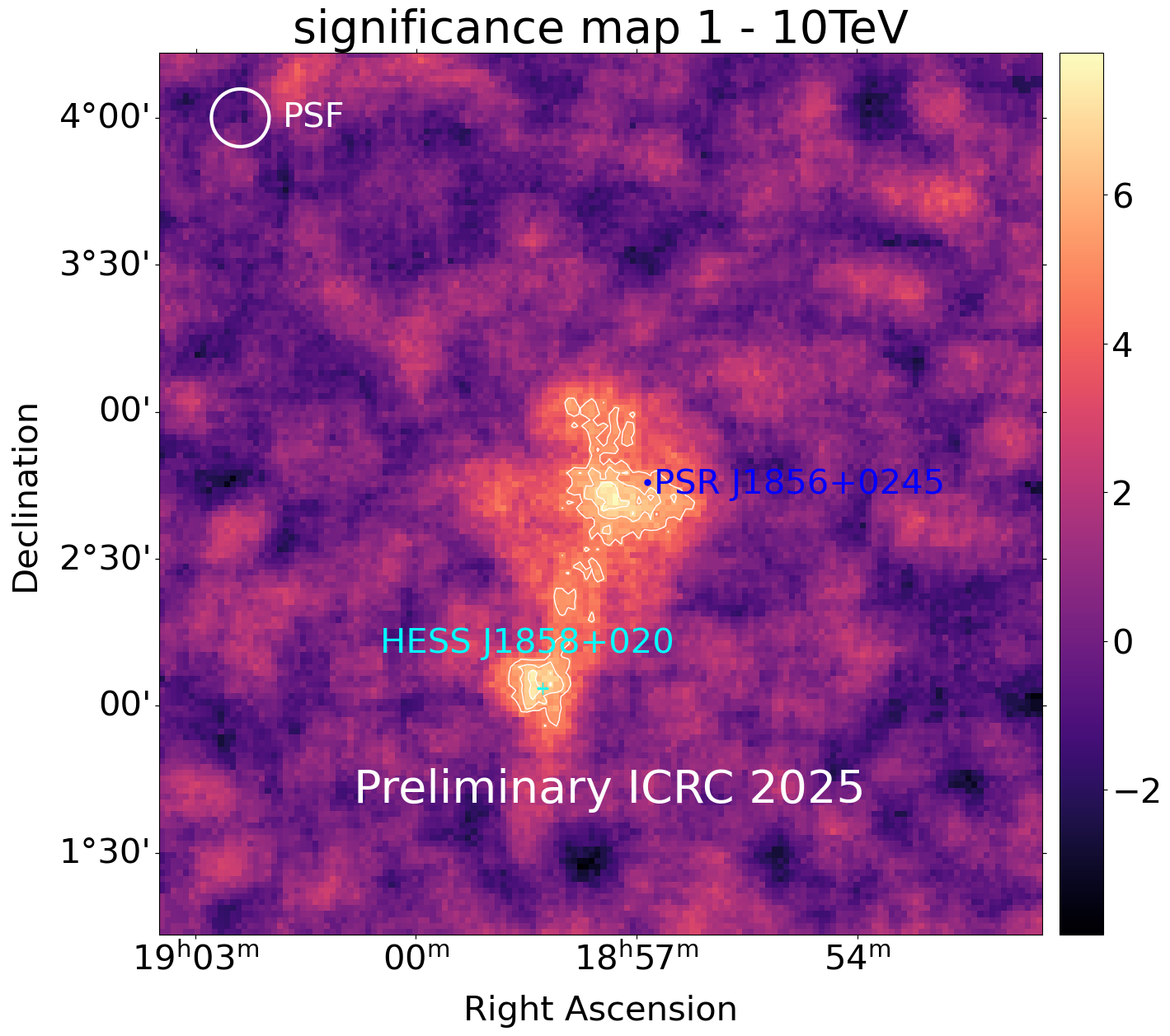}
    \caption{Significance map of region around HESS~J1857+026 in 0.3\,-\,1\,TeV (left) and in 1\,-\,10\,TeV (right). The white contours represent significance values of 5, 6, and 7\,$\sigma$. The blue dot marks the location of PSR~J1856+0245.}
    \label{fig:significance_map}
\end{figure}

After the optimized model of HESS~J1857+026 is determined, the morphological features of the source are characterized by constructing its radial profile which is the flux per solid angle as a function of the offset from the TeV center. First, the sky region is divided with annuli of equal widths using the best-fit R.A. and Dec. of the Gaussian centroid in the spatial model as the center. Gammapy's FluxProfileEstimator is then employed where the spectral index of the spectral model is also used as an input to estimate the total flux within each annulus. Finally, the flux of each annulus is divided by the solid angle it subtends. The radial profile describes how fast the VHE emission decays as the distance to the center increases, which can be used to infer the diffusion speed of the underlying electron population, as described in Section \ref{sec:diffusion}.

\section{VERITAS analysis results}\label{sec:results}
\subsection{Significance map and energy spectra}
The test statistic of the best-fit model for HESS~J1857+026 corresponds to 15.3 $\sigma$ above the null hypothesis. Fig. \ref{fig:significance_map} shows the significance map of the region around HESS~J1857+026 in 0.3\,-\,1\,TeV and in 1\,-\,10\,TeV. PSR~J1856+0245 is clearly displaced from the VHE emission center, confirming the findings by previous studies \cite{hessels_psr_2008,magic_2014}. A northern component shows up in energies above 1 TeV. This additional structure could indicate a separate source proposed by MAGIC in \cite{magic_2014} or it could originate from the expanse of the source itself due to faster diffusion of electrons with higher energies. However, more data are needed to make any further conclusions.

\begin{figure}[h]
    \centering
    \includegraphics[width=0.8\linewidth]{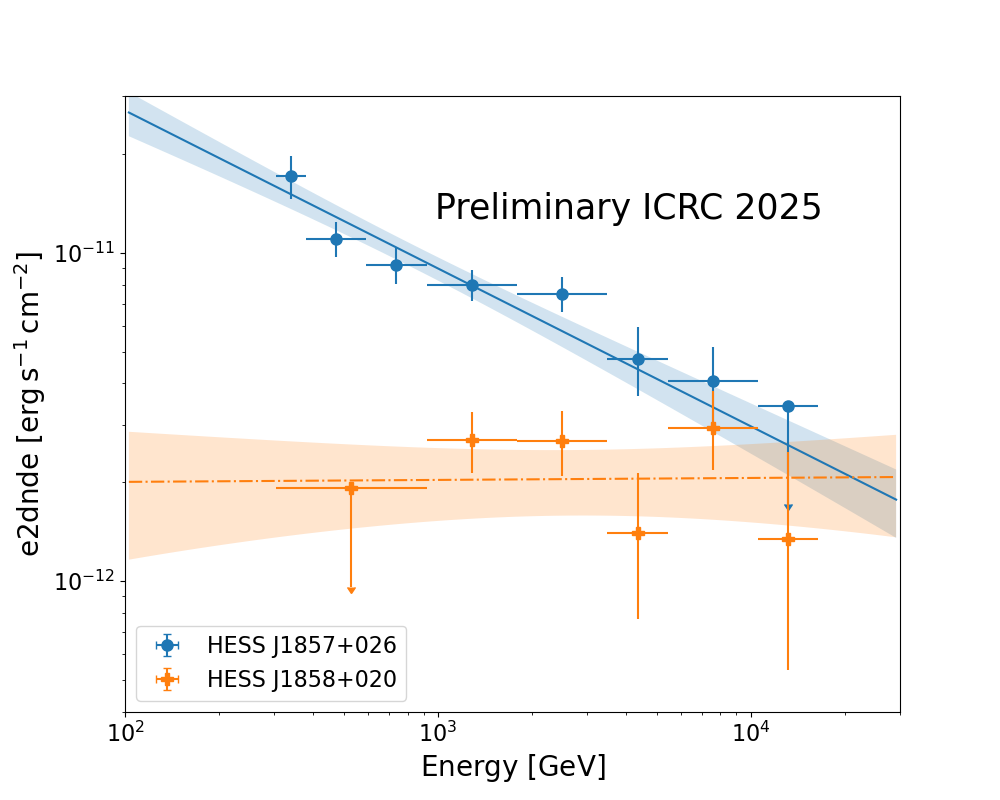}
    \caption{Spectra of HESS~J1857+026 and HESS~J1858+020. The spectral points are extracted from the stacked dataset using the spectral index determined from the fit for each source. The lines describe the best-fit power-law spectra of the models.}
    \label{fig:sed}
\end{figure}
The spectra of HESS~J1857+026 and HESS~J1858+020 of the optimized model are shown in Fig. \ref{fig:sed}. Both sources can be well described by the a power law below 10 TeV. We note the upper limit set by VERITAS beyond 10 TeV. Spectral feature of this source at the highest energy will be discussed in a HAWC contribution in this conference as well as in the upcoming paper. The best-fit free parameters of the model are listed in table \ref{tab:VTS_gammapy_model_fit}.

\begin{table}[h]
\centering
    \begin{tabular}{cccccc}
        \hline
        \hline
              & HESS J1857+026 & HESS J1858+020 \\
        \hline
        R.A. ($^{\circ}$)  & 284.37 $\pm$ 0.017 & 284.58 $\pm$ 0.023 \\
        Dec. ($^{\circ}$)  & 2.72   $\pm$ 0.020 & 2.06   $\pm$ 0.029 \\
        Gaussian extension ($^{\circ}$) & 0.24 $\pm$ 0.015  & 0.13   $\pm$  0.037\\
        N$_{0}$ (TeV$^{-1}$s$^{-1}$cm$^{-2}$) &  (5.5 $\pm$ 0.43$)\times 10^{-12}$ & (1.2 $\pm$ 0.30$)\times 10^{-12}$\\
        $\Gamma$ & 2.46 $\pm$ 0.060 & 2.0 $\pm$ 0.11\\
        \hline
    \end{tabular}
\caption{Optimized parameters for HESS~J1857+026 and HESS~J1858+020 in the VERITAS 0.3--10\,TeV source model.}
\label{tab:VTS_gammapy_model_fit}
\end{table}

\subsection{Diffusion coefficient}\label{sec:diffusion}
\begin{figure}[h]
    \centering
    \includegraphics[width=0.8\linewidth]{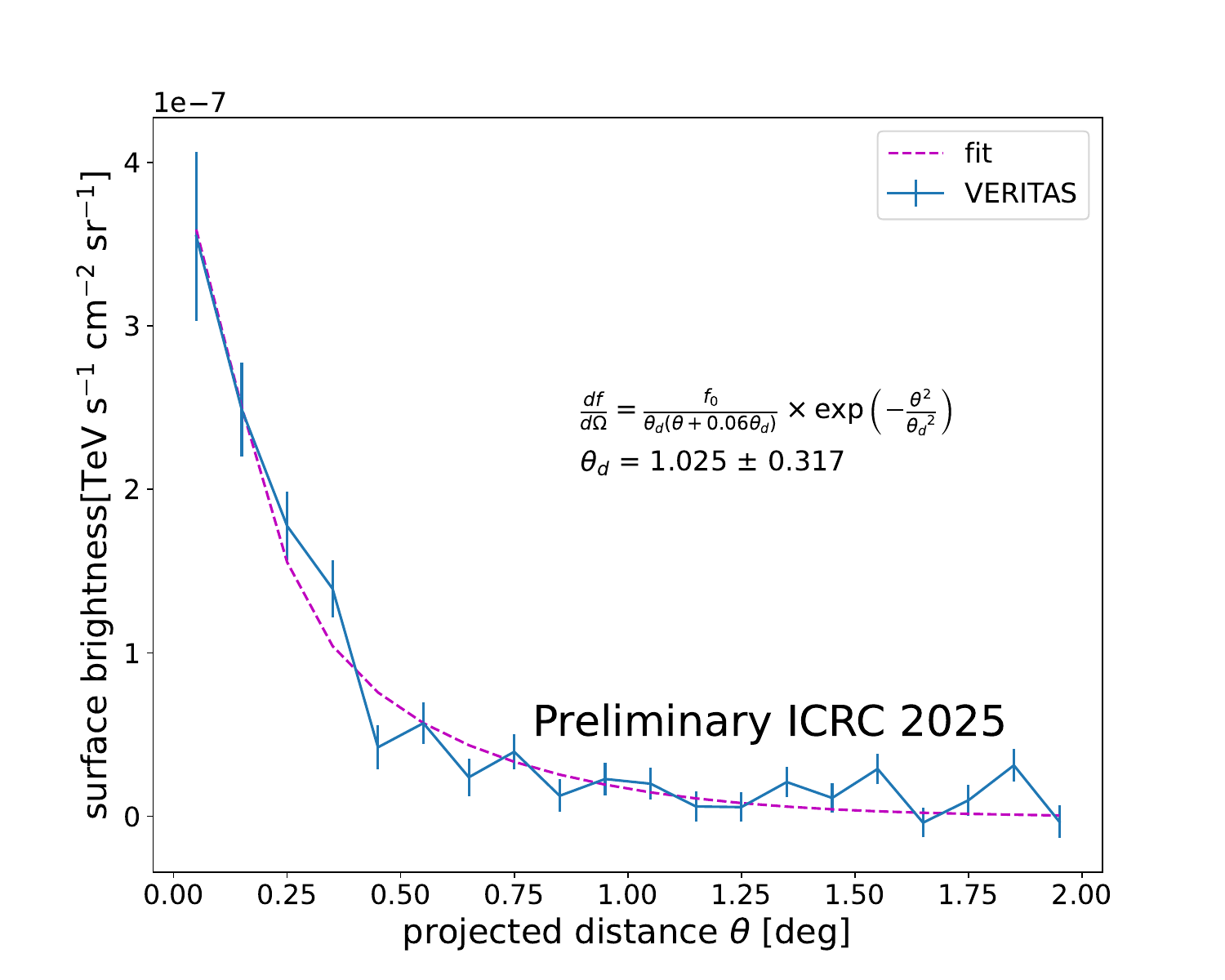}
    \caption{Radial profile of HESS~J1857+026 in the energy range of 0.3\,-\,10\,TeV. Dashed magenta line is the fit according to Eq. \ref{eq:radial} and the best-fit paramter $\theta_d$ is printed in the figure.}
    \label{fig:radial}
\end{figure}

Fig. \ref{fig:radial} shows the radial profile of HESS~J1857+026 in 0.3\,-\,10\,TeV constructed by following the procedures explained in Section \ref{sec:VTS-analysis}. If an electron population were responsible for the observed VHE flux through inverse Compton scattering of the background photons and if they were continuously injected at a constant rate and diffused away from the center, the flux per solid angle as a function of offset $\theta$ can be described with equation \cite{HAWC_2017_Geminga}
\begin{equation}\label{eq:radial}
    \frac{df}{d\Omega} = \frac{f_0}{\theta_d(\theta+0.06\theta_d)}\times \exp \left(-\frac{\theta^2}{\theta_d^2} \right).
\end{equation}
Here $f_0$ is the normalization factor. $\theta_d$ characterizes how fast the diffusion process is. A faster diffusion of electrons leads to a larger tail while making the center region dimmer and harder to detect. A fit on the radial profile with Eq. \ref{eq:radial} results in a $\theta_d$ value of  $1.025\pm0.317\,^\circ$. The diffusion length $d$ is calculated as $d=\theta_d\times l=98.4 \pm 30.4 $\,pc where $l$ is the distance to the pulsar taken as 5.5\,kpc from \cite{petriella_radio_2021}. $d$ is also related to the diffusion coefficient $D$ through
\begin{equation}
    d=2\sqrt{Dt}
\end{equation}
where t is the available time for the diffusion process. If one assumes that the magnetic field is on the order of $\sim$1 $\mu$G, a quick calculation shows that cooling time for the electron population responsible for the gamma rays observed by VERITAS are on the order of tens of kyr, larger or comparable to the age of the pulsar. Therefore, we estimate the diffusion by assuming that t $\sim$ 21\,kyr \cite{hessels_psr_2008}, and the diffusion coefficient is calculated as $(3.5 \pm 2.2)\times 10^{28}$\,cm$^2$/s, which is still an order of magnitude lower than the Galactic average noted in \cite{HAWC_2017_Geminga,genolini_cosmic-ray_2019}.

\section{Summary}
VERITAS has significantly detected HESS~J1857+026. The morphology of the source seems to indicate an expansion of the source region or an unrelated source with increased energy. Nevertheless, the diffusion coefficient of the electrons under a leptonic scenario is calculated and is shown to be an order of magnitude below the Galactic average. A future multi-wavelength paper is being prepared to fully characterize the emission mechanism of HESS~J1857+026 and a more careful consideration of diffusion will also be presented.

\acknowledgments{This research is supported by grants from the U.S. Department of Energy Office of Science, the U.S. National Science Foundation and the Smithsonian Institution, by NSERC in Canada, and by the Helmholtz Association in Germany. This research used resources provided by the Open Science Grid, which is supported by the National Science Foundation and the U.S. Department of Energy's Office of Science, and resources of the National Energy Research Scientific Computing Center (NERSC), a U.S. Department of Energy Office of Science User Facility operated under Contract No. DE-AC02-05CH11231. We acknowledge the excellent work of the technical support staff at the Fred Lawrence Whipple Observatory and at the collaborating institutions in the construction and operation of the instrument.}
\bibliographystyle{JHEP}
\bibliography{mybib}

\providecommand{\href}[2]{#2}\begingroup\raggedright\begin{thebibliography}{10}

\bibitem{hess_2008}
F.~Aharonian, A.~G. Akhperjanian, U.~Barres De~Almeida, A.~R. Bazer-Bachi, B.~Behera, M.~Beilicke et~al., \emph{{HESS} very-high-energy gamma-ray sources without identified counterparts}, \href{https://doi.org/10.1051/0004-6361:20078516}{\emph{Astronomy \& Astrophysics} {\bfseries 477} (2008) 353}.

\bibitem{hessels_psr_2008}
J.~W.~T. Hessels, D.~J. Nice, B.~M. Gaensler, V.~M. Kaspi, D.~R. Lorimer, D.~J. Champion et~al., \emph{{PSR} {J1856}+0245: {Arecibo} {Discovery} of a {Young}, {Energetic} {Pulsar} {Coincident} with the {TeV} $\gamma$-{Ray} {Source} {HESS} {J1857}+026}, \href{https://doi.org/10.1086/590908}{\emph{The Astrophysical Journal} {\bfseries 682} (2008) L41}.

\bibitem{Fermi_2017_extsources}
M.~Ackermann, M.~Ajello, L.~Baldini, J.~Ballet, G.~Barbiellini, D.~Bastieri et~al., \emph{Search for {Extended} {Sources} in the {Galactic} {Plane} {Using} {Six} {Years} of {Fermi}-{Large} {Area} {Telescope} {Pass} 8 {Data} above 10 {GeV}}, \href{https://doi.org/10.3847/1538-4357/aa775a}{\emph{The Astrophysical Journal} {\bfseries 843} (2017) 139}.

\bibitem{magic_2014}
J.~Aleksić, S.~Ansoldi, L.~A. Antonelli, P.~Antoranz, A.~Babic, P.~Bangale et~al., \emph{{MAGIC} reveals a complex morphology within the unidentified gamma-ray source {HESS} {J1857}+026}, \href{https://doi.org/10.1051/0004-6361/201423517}{\emph{Astronomy \& Astrophysics} {\bfseries 571} (2014) A96}.

\bibitem{petriella_radio_2021}
A.~Petriella, L.~Duvidovich and E.~Giacani, \emph{Radio study of {HESS} {J1857}+026: {Gamma}-rays from a superbubble?}, \href{https://doi.org/10.1051/0004-6361/202141254}{\emph{Astronomy \& Astrophysics} {\bfseries 652} (2021) A142}.

\bibitem{HAWC_2017_Geminga}
A.~U. Abeysekara, A.~Albert, R.~Alfaro, C.~Alvarez, J.~D. Álvarez, R.~Arceo et~al., \emph{Extended gamma-ray sources around pulsars constrain the origin of the positron flux at {Earth}}, \href{https://doi.org/10.1126/science.aan4880}{\emph{Science} {\bfseries 358} (2017) 911}.

\bibitem{genolini_cosmic-ray_2019}
Y.~Genolini, M.~Boudaud, P.~I. Batista, S.~Caroff, L.~Derome, J.~Lavalle et~al., \emph{Cosmic-ray transport from {AMS}-02 {B}/{C} data: benchmark models and interpretation}, \href{https://doi.org/10.1103/PhysRevD.99.123028}{\emph{Physical Review D} {\bfseries 99} (2019) 123028}.

\bibitem{di_mauro_evidences_2020}
M.~Di~Mauro, S.~Manconi and F.~Donato, \emph{Evidences of low-diffusion bubbles around {Galactic} pulsars}, \href{https://doi.org/10.1103/PhysRevD.101.103035}{\emph{Physical Review D} {\bfseries 101} (2020) 103035}.

\bibitem{HAWC_2025_halo}
A.~Albert, R.~Alfaro, C.~Alvarez, J.~C. Arteaga-Vel\'azquez, D.~Avila~Rojas, H.~A. Ayala~Solares et~al., \emph{Extended {TeV} halos may commonly exist around middle-aged pulsars}, \href{https://doi.org/10.1103/PhysRevLett.134.171005}{\emph{Phys. Rev. Lett.} {\bfseries 134} (2025) 171005}.

\bibitem{hess_2018_GPS}
H.~Abdalla, A.~Abramowski, F.~Aharonian, F.~A. Benkhali, E.~O. Angüner, M.~Arakawa et~al., \emph{The {H}.{E}.{S}.{S}. {Galactic} plane survey}, \href{https://doi.org/10.1051/0004-6361/201732098}{\emph{Astronomy \& Astrophysics} {\bfseries 612} (2018) A1}.

\bibitem{maier2017eventdisplay}
G.~Maier and J.~Holder, \emph{Eventdisplay: an analysis and reconstruction package for ground-based gamma-ray astronomy}, {\emph{arXiv preprint arXiv:1708.04048} (2017) }.

\bibitem{Maier_Eventdisplay_An_Analysis_2024}
G.~Maier, J.~Holder, A.~McCann, B.~Behera, C.~Duke, C.~Giuri et~al., \emph{{Eventdisplay: An Analysis and Reconstruction Package for Ground-based Gamma-ray Astronomy}},  Dec., 2024.
\newblock 10.5281/zenodo.3559075.

\bibitem{gammapy_zenodo_v1p3}
F.~Acero, A.~Aguasca-Cabot, J.~Bernete, N.~Biederbeck, J.~Djuvsland, A.~Donath et~al., \emph{Gammapy: Python toolbox for gamma-ray astronomy},  Jan., 2025.
\newblock 10.5281/zenodo.14760974.

\bibitem{gammapy:2023}
A.~{Donath}, R.~{Terrier}, Q.~{Remy}, A.~{Sinha}, C.~{Nigro}, F.~{Pintore} et~al., \emph{Gammapy: A python package for gamma-ray astronomy}, \href{https://doi.org/10.1051/0004-6361/202346488}{\emph{Astronomy \& Astrophysics} {\bfseries 678} (2023) A157}.

\end{thebibliography}\endgroup

\section*{Full Author List: VERITAS Collaboration}

\scriptsize
\noindent
A.~Archer$^{1}$,
P.~Bangale$^{2}$,
J.~T.~Bartkoske$^{3}$,
W.~Benbow$^{4}$,
Y.~Chen$^{5}$,
J.~L.~Christiansen$^{6}$,
A.~J.~Chromey$^{4}$,
A.~Duerr$^{3}$,
M.~Errando$^{7}$,
M.~Escobar~Godoy$^{8}$,
J.~Escudero Pedrosa$^{4}$,
Q.~Feng$^{3}$,
S.~Filbert$^{3}$,
L.~Fortson$^{9}$,
A.~Furniss$^{8}$,
W.~Hanlon$^{4}$,
O.~Hervet$^{8}$,
C.~E.~Hinrichs$^{4,10}$,
J.~Holder$^{11}$,
T.~B.~Humensky$^{12,13}$,
M.~Iskakova$^{7}$,
W.~Jin$^{5}$,
M.~N.~Johnson$^{8}$,
E.~Joshi$^{14}$,
M.~Kertzman$^{1}$,
M.~Kherlakian$^{15}$,
D.~Kieda$^{3}$,
T.~K.~Kleiner$^{14}$,
N.~Korzoun$^{11}$,
S.~Kumar$^{12}$,
M.~J.~Lang$^{16}$,
M.~Lundy$^{17}$,
G.~Maier$^{14}$,
C.~E~McGrath$^{18}$,
P.~Moriarty$^{16}$,
R.~Mukherjee$^{19}$,
W.~Ning$^{5}$,
R.~A.~Ong$^{5}$,
A.~Pandey$^{3}$,
M.~Pohl$^{20,14}$,
E.~Pueschel$^{15}$,
J.~Quinn$^{18}$,
P.~L.~Rabinowitz$^{7}$,
K.~Ragan$^{17}$,
P.~T.~Reynolds$^{21}$,
D.~Ribeiro$^{9}$,
E.~Roache$^{4}$,
I.~Sadeh$^{14}$,
L.~Saha$^{4}$,
H.~Salzmann$^{8}$,
M.~Santander$^{22}$,
G.~H.~Sembroski$^{23}$,
B.~Shen$^{12}$,
M.~Splettstoesser$^{8}$,
A.~K.~Talluri$^{9}$,
S.~Tandon$^{19}$,
J.~V.~Tucci$^{24}$,
J.~Valverde$^{25,13}$,
V.~V.~Vassiliev$^{5}$,
D.~A.~Williams$^{8}$,
S.~L.~Wong$^{17}$,
T.~Yoshikoshi$^{26}$\\
\\
\noindent
$^{1}${Department of Physics and Astronomy, DePauw University, Greencastle, IN 46135-0037, USA}

\noindent
$^{2}${Department of Physics, Temple University, Philadelphia, PA 19122, USA}

\noindent
$^{3}${Department of Physics and Astronomy, University of Utah, Salt Lake City, UT 84112, USA}

\noindent
$^{4}${Center for Astrophysics $|$ Harvard \& Smithsonian, Cambridge, MA 02138, USA}

\noindent
$^{5}${Department of Physics and Astronomy, University of California, Los Angeles, CA 90095, USA}

\noindent
$^{6}${Physics Department, California Polytechnic State University, San Luis Obispo, CA 94307, USA}

\noindent
$^{7}${Department of Physics, Washington University, St. Louis, MO 63130, USA}

\noindent
$^{8}${Santa Cruz Institute for Particle Physics and Department of Physics, University of California, Santa Cruz, CA 95064, USA}

\noindent
$^{9}${School of Physics and Astronomy, University of Minnesota, Minneapolis, MN 55455, USA}

\noindent
$^{10}${Department of Physics and Astronomy, Dartmouth College, 6127 Wilder Laboratory, Hanover, NH 03755 USA}

\noindent
$^{11}${Department of Physics and Astronomy and the Bartol Research Institute, University of Delaware, Newark, DE 19716, USA}

\noindent
$^{12}${Department of Physics, University of Maryland, College Park, MD, USA }

\noindent
$^{13}${NASA GSFC, Greenbelt, MD 20771, USA}

\noindent
$^{14}${DESY, Platanenallee 6, 15738 Zeuthen, Germany}

\noindent
$^{15}${Fakult\"at f\"ur Physik \& Astronomie, Ruhr-Universit\"at Bochum, D-44780 Bochum, Germany}

\noindent
$^{16}${School of Natural Sciences, University of Galway, University Road, Galway, H91 TK33, Ireland}

\noindent
$^{17}${Physics Department, McGill University, Montreal, QC H3A 2T8, Canada}

\noindent
$^{18}${School of Physics, University College Dublin, Belfield, Dublin 4, Ireland}

\noindent
$^{19}${Department of Physics and Astronomy, Barnard College, Columbia University, NY 10027, USA}

\noindent
$^{20}${Institute of Physics and Astronomy, University of Potsdam, 14476 Potsdam-Golm, Germany}

\noindent
$^{21}${Department of Physical Sciences, Munster Technological University, Bishopstown, Cork, T12 P928, Ireland}

\noindent
$^{22}${Department of Physics and Astronomy, University of Alabama, Tuscaloosa, AL 35487, USA}

\noindent
$^{23}${Department of Physics and Astronomy, Purdue University, West Lafayette, IN 47907, USA}

\noindent
$^{24}${Department of Physics, Indiana University Indianapolis, Indianapolis, Indiana 46202, USA}

\noindent
$^{25}${Department of Physics, University of Maryland, Baltimore County, Baltimore MD 21250, USA}

\noindent
$^{26}${Institute for Cosmic Ray Research, University of Tokyo, 5-1-5, Kashiwa-no-ha, Kashiwa, Chiba 277-8582, Japan}

\end{document}